# Robust curvelet domain watermarking technique that preserves cleanness of high quality images

Wook-Hyung Kim, Seung-Hun Nam, Ji-Hyeon Kang, and Heung-Kyu Lee

**Abstract**

Watermarking inserts invisible data into content to protect copyright. The embedded information provides proof of authorship and facilitates tracking illegal distribution, etc. Current robust watermarking techniques have been proposed to preserve inserted copyright information from various attacks, such as content modification and watermark removal attack. However, since the watermark is inserted in the form of noise, there is an inevitable effect of reducing content visual quality. In general, more robust watermarking techniques tend to have larger effect on the quality, and content creators and users are often reluctant to insert watermarks. Thus, there is a demand for a watermark that maintains maximum image quality, even if the watermark performance is slightly inferior. Therefore, we propose a watermarking technique that maximizes invisibility while maintaining sufficient robustness and data capacity enough to be applied for real situations. The proposed method minimizes watermarking energy by adopting curvelet domain multi-directional decomposition to maximize invisibility, and maximizes robustness against signal processing attack by watermarking pattern suitable for curvelet transformation. The method is also robust against geometric attack by employing watermark detection method utilizing curvelet characteristics. The proposed method showed very good results of 57.65 dB peak signal-to-noise ratio in fidelity tests, and mean opinion score showed that images treated with the proposed method were hardly distinguishable from the originals. The proposed technique also showed good robustness against signal processing and geometric attacks compared with existing techniques.

**Index Terms**

Content copyright protection; Digital content watermark; Curvelet transform; High quality content; Blind detection

## I. INTRODUCTION

Watermarking has emerged as one method to prevent copyright infringement. Invisible copyright information is inserted into the content, as noise, so it is not easily noticeable. However, because of the noise form, the watermark degrades content quality. In particular, watermarking methods that are robust against various attacks can significantly degrade image quality, due to the large watermark embedding energy. Figure 1 shows that the watermarked image (Fig. 1(b)), is visually compromised compared to the original image (Fig. 1(a)). Any reader who can distinguish small image changes, and the actual content producers, would notice this level of degradation, and content producers and users of high quality content are reluctant to insert watermarks in images. High resolution and high quality images, such as ultra high definition (UHD), have become popular, and image quality has become more important. Consequently, there has been high demand for watermarking technology focusing on image quality rather than robustness and data capacity.

The proposed method maximizes invisibility by adopting the curvelet domain [1] for watermark embedding. The curvelet transform can decompose an image in more than 8 directions, depending on the domain configuration, so is advantageous to insert a watermark of smaller energy. Several studies have considered the curvelet domain previously.

Zhang et al. [2] proposed a method to embed and extract watermarks in the amplitude of curvelet coefficients using quantization index modulation (QIM) [3]. The method was able to detect watermarks blindly, and was robust against various filter, compression, and noise attacks when the embedded watermark energy was large. However, the approach did not consider curvelet filter characteristics to cut frequency components in a specific direction during curvelet transform, hence detection rate was somewhat lower than the embedded watermark energy.

Tao et al. [4] proposed a method to embedding watermarks into the curvelet coefficients using the spread spectrum [5]. The method was capable of blind detection and was robust to signal distortion. However, it also failed to consider curvelet filter characteristics, and hence also had lower detection rate than the watermark embedding energy, and was vulnerable to geometric attack, such as image scaling and rotation.

Channapragada et al. [6] proposed a curvelet watermarking method using magic squares. This method resized the watermark to the same as the image using the magic square method [7], and embedded the resized watermark into the curvelet image using the spread spectrum. The resultant watermark had excellent invisibility and robustness to various attacks, but was impractical because it is a non-blind method that required the original image to detect the watermark.

This paper proposes a watermarking method that maximizes invisibility while maintaining robustness against attacks that occur frequently in real conditions. To achieve this, we adopted a curvelet domain to minimize watermark embedding energy. However, due to inherent curvelet filter characteristics, watermark signals are distorted in the forward and inverse curvelet transformation processes when a watermark is embedded with conventional watermarking methods. To prevent this, we adopt a particular pattern generation method suitable for curvelets. We also present robust detection methods and templates for geometric attacks. The proposed method achieves the following contributions: 1. High invisibility that does not significantly impair image quality. 2. Blind watermarking, i.e., does not require the original image for watermark detection. 3. Robustness against various signal attacks with low watermarking energy. 4. Robustness against geometric attacks, such as scaling and rotation. The remainder of this paper is organized as follows. Section2 provides a brief introduction of the curvelet transform, and Section 3 discusses the proposed watermarking algorithm. Sections 4 present experimental results and Section 5 concludes the paper.

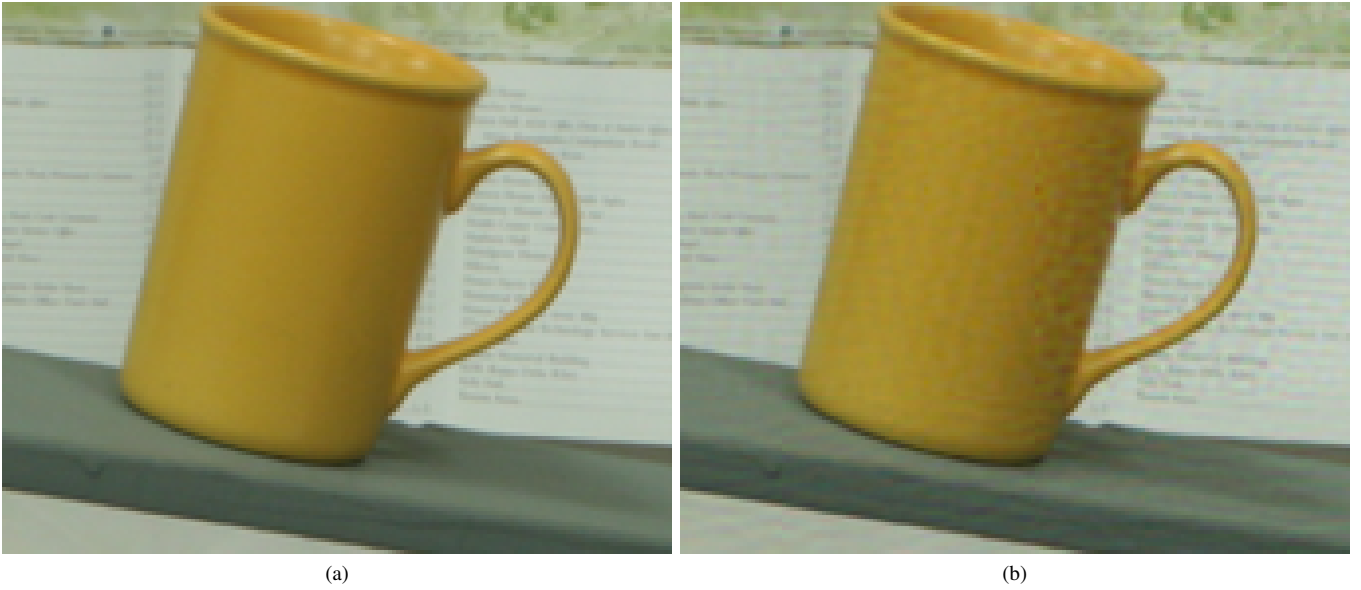

Fig. 1: Image quality degradation due to watermark embedding: enlarged (a) original and (b) watermarked image.

## II. CURVELET TRANSFROM

In contrast to conventional domain watermarking methods, curvelet domain watermarks are distorted during forward and inverse curvelet transform. This section provides a brief description of the curvelet domain and explains why the watermark is corrupted during the curvelet transform.

### A. A Brief Overview of Curvelet Transform

The curvelet transform is a multi-scale decomposition-like wavelet transform, and the curvelet represents the curve shape for the various directions in the spatial domain [8]–[11]. The curvelet transform is developed to improve on the limitation of wavelet-based transforms and can represent edges more efficiently than conventional wavelet-based transforms. Moreover, curvelet bases cover all frequencies in contrast to other directional multi-scale transforms, such as the Gabor and Ridgelet transforms [12]. The curvelet transform is expressed as follows:

$$C(g,l,k) := \langle f, \varphi_{g,l,k} \rangle = \int_{\mathbf{R}^2} f(x)\overline{\varphi_{g,l,k}} dx \\ = \frac{1}{(2\pi)^2} \int \hat{f}(\omega) U_g(R_{\theta_l}\omega) e^{i\langle x_k^{(g,l)}, \omega \rangle} d\omega, \quad (1)$$

$$U_g(r,\theta) = 2^{-3g/4} W(2^{-g}r) V\Big(\frac{2^{\lfloor g/2 \rfloor}\theta}{2\pi}\Big), \quad (2)$$

In (1), $C$ is the curvelet coefficient, $g = 0, 1, 2, ...$ is the scale parameter, $l$ is the rotation parameter, and $k = (k_1, k_2)$ is the translation parameter. $U_g$ is a "wedge"-shaped frequency window represented in (2). $R_\theta$ is the rotation operator and $\theta_l = 2\pi \cdot 2^{-\lfloor g/2 \rfloor} \cdot l$. In (2), $W$ and $V$ are the radial and angular windows, respectively.

The curvelet is illustrated in Fig. 2. Fig. 2 (a) illustrates the tiling of the curvelet in the frequency domain, and the curvelet shape in several directions and scales in the spatial domain are shown in Fig. 2 (b)–(d). The frequency is divided into various directions and various scales, which simplifies minimizing watermark embedding energy.

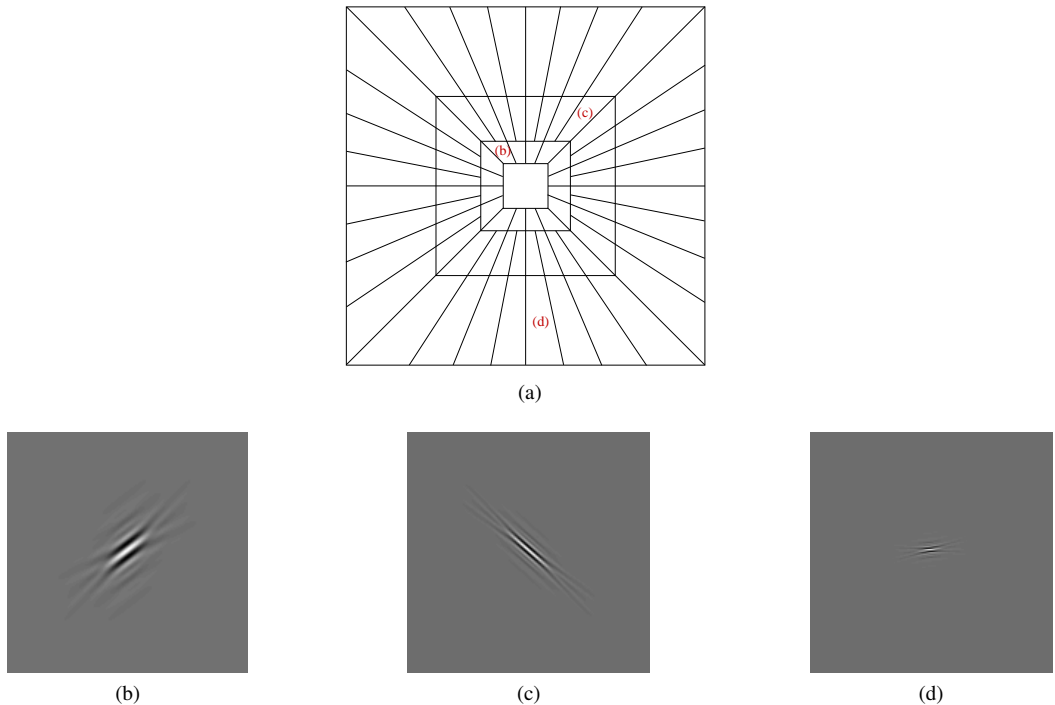

Fig. 2: Curvelet in the frequency and spatial domain. (a) Curvelet tiling of the frequency domain; (b)–(d) Curvelets for various scales and directions in the spatial domain. Curvelets are drawn on $k_1 = w/2$ and $k_2 = h/2$.

### B. Problem of Watermarking on the Curvelet Domain

Fig. 3 shows a diagram of forward curvelet transform. The inverse transform is similar to the forward transform, and the image passes through the curvelet filter in both the forward and inverse transform. The curvelet filter consists of frequency components in a specific direction, as shown in Fig. 4 (a). On the other hand, watermarks embedded by spread spectrum and QIM include all the frequency components, as shown in Figs. 4 (b) and 4 (c). The inserted watermark passes through the filter during the curvelet transform, and the frequency components outside the filter are removed. This causes the embedded watermark in the curvelet image to be corrupted during transformation, which reduces detection rate. A watermarking technique specifically for the curvelet domain is required to prevent this corruption.

## III. PROPOSED WATERMARKING METHOD IN CURVELET DOMAIN

This section describes the proposed watermarking algorithm. Fig 5 shows the proposed embedding and detection process. We designed a watermark pattern that is not damaged during curvelet transformation, and watermark embedding and detection were performed using this pattern.

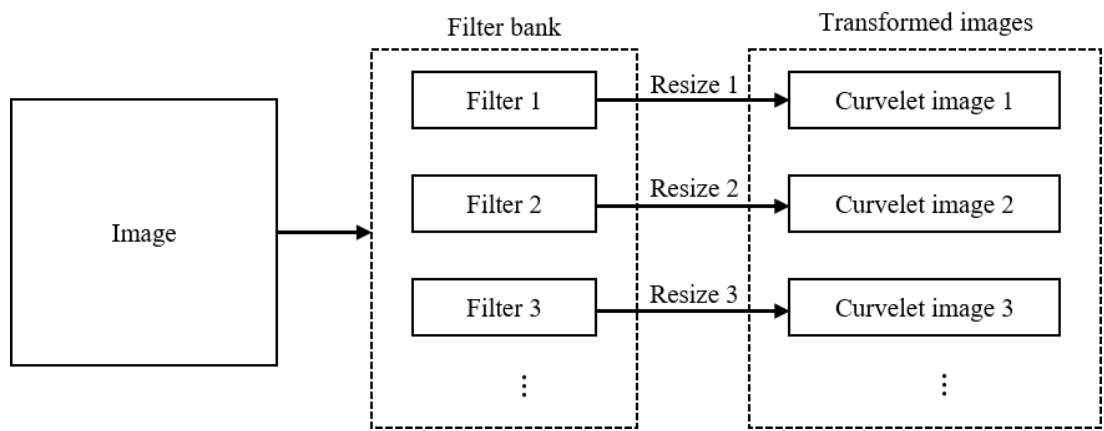

Fig. 3: Diagram of curvelet transform.

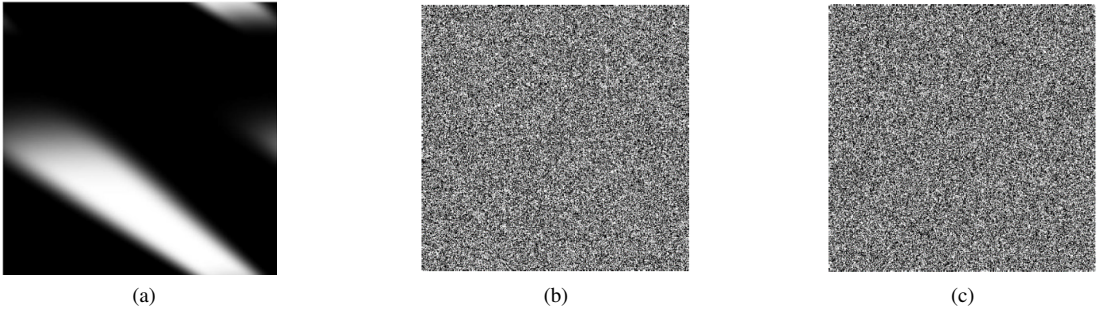

Fig. 4: Frequency components of (a) Curvelet filter, (b) Spread spectrum watermark, (c) quantization watermark (scale =3 and direction = 1).

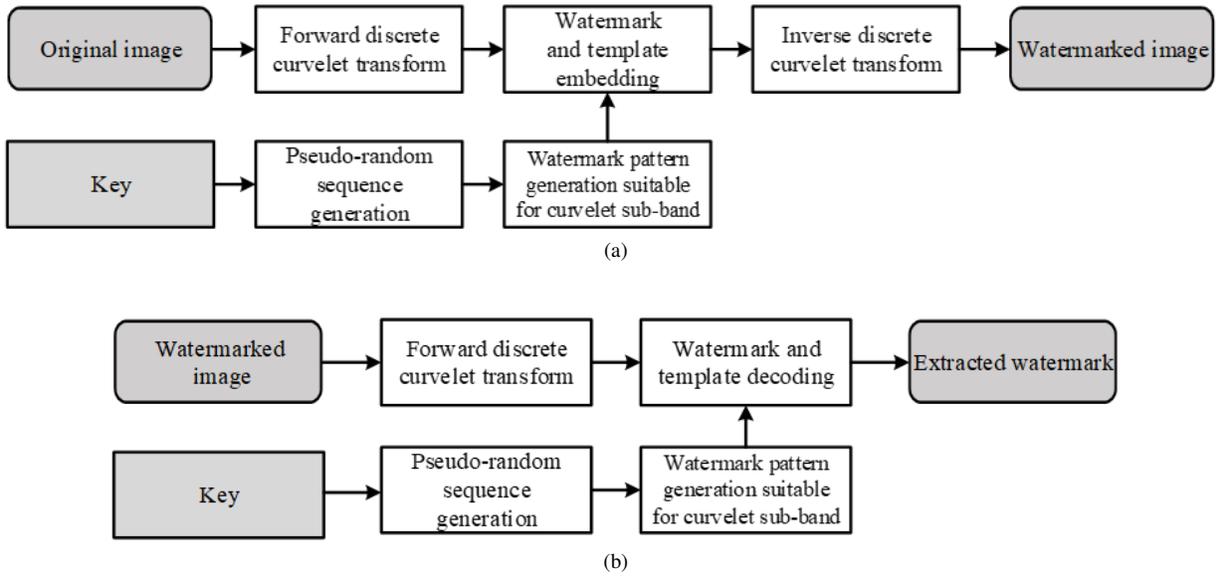

Fig. 5: Proposed curvelet domain watermarking method: (a) embedding and (b) Detection procedures.

### A. Watermark Pattern Design for the Curvelet Domain

To address the problems discussed in Section II, we adopt a watermark pattern that passes through curvelet filtering without distortion. To avoid confusion, $\bar{S}$ is defined as the spatial domain, $\bar{T}$ is defined as the frequency domain of spatial domain, and $\bar{C}$ is defined as the curvelet domain. $\bar{C}$ is composed of frequency and spatial components, but when discrete Fourier transform (DFT) is applied, the transformed domain, $\bar{F}$, only includes frequency components. The symbols are summarized in Table I.

To pass through the curvelet filter without damage, the watermark pattern must be designed using only internal frequencies of the curvelet filter. We present two methods to design such a watermark pattern.

1. **Simultaneous equation.** We solved the simultaneous equation to obtain a watermark pattern incorporating only frequency components inside the curvelet filter,

$$\sum_{(u,v) \in \boldsymbol{A}} k_{u,v} \cdot \boldsymbol{F}_{u,v} = \boldsymbol{W}, \tag{3}$$

where $k$ is the DFT coefficient in the $\bar{F}$ domain, $\boldsymbol{F}$ is the inverse DFT matrix from the $\bar{F}$ to the $\bar{C}$ domain, $(u,v)$ is the coordinate of the $\bar{F}$ domain, and $\boldsymbol{A}$ is the set of coordinates inside the curvelet filter on $\bar{F}$ (i.e., the bright part of Fig. 4 (a)). Equation 3 is the same as inverse discrete Fourier transform (IDFT), but uses limited frequency components. Since this simultaneous equation is overdetermined, there is often no solution, so we find a solution $\tilde{W}$ that is close to $W$. This method can insert a watermark in a desired position for a desired embedding method (such as spread spectrum or QIM), but has a disadvantage of requiring significant computational overhead. To obtain the watermark pattern following this method, several thousand-dimensional simultaneous equation must be solved for full high definition image.

TABLE I: Domain symbol definitions. The frequency domain of curvelet domain is the DFT of $\bar{C}$

| | Spatial domain | Frequency domain of spatial domain | Curvelet domain | Frequency domain of curvelet domain |
|---|---|---|---|---|
| Symbol | $S$ | $T$ | $C$ | $F$ |

2. **Random sequence.** A random sequence is scattered inside the filter of the $barF$ domain and the pattern is obtained by applying IDFT to the scattered random sequence. First, a random sequence is generated, equal length to the number of coordinates in the curvelet filter (i.e., the number of elements in $\boldsymbol{A}$). The generated sequence is then substituted into the curvelet filter in order. Finally, applying IDFT to the sequences generates a watermark pattern that is not corrupted by the curvelet filter. Since the mean value of the generated watermark pattern is approximately 0, only the variance needs to be amplified to 1. This method has the disadvantage of only inserting a watermark using the spread spectrum method and cannot select the watermark position, but it has the advantage of requiring relatively little computation.

The first method is impractical due to high computational complexity. It is also necessary to solve additional problems such as finding an optimal $\tilde{W}$ similar to $W$ in order to minimize the watermark signal being filtered. Therefore, we uses the second method for simplicity and practicality.

## B. Embedding Method

Figure 5(a) shows the watermark embedding process. The original image is transformed into the curvelet domain. A random sequence generated using the key, and the watermark pattern is generated as described in Section III-A. The generated watermark pattern is then inserted into the curvelet image using the spread spectrum method [13]. The process can be represented as

$$C'_{s,d}(m,n) = C_{s,d}(m,n) + \alpha |C_{s,d}(m,n)| W_{s,d}(m,n), \quad (4)$$

where $1 \leq m \leq i, 1 \leq n \leq j$; $C$ is the curvelet coefficient of the original; $C'$ is a watermarked curvelet coefficient; $s$ and $d$ are the scale and direction, respectively, that the watermark is to be inserted; $m$ and $n$ are the horizontal and vertical coordinates, respectively, of the curvelet domain; $W$ is the watermark; $i$ and $j$ are the horizontal and vertical size, respectively, of the curvelet image, and $\alpha$ is the watermark embedding strength.

Equation 4 is for a single scale and direction, and it is possible to embed multiple watermarks by repeating Eq. 4 for various scales and directions. We also embed the template in the other direction, in the same way as the watermark, as shown in Algorithm 1. Algorithm 1 describes a situation where a watermark is inserted into scale 3 and direction 1, and a template is inserted into scale 3 and direction 9. This provides robustness against rotation attacks and explains in detail the role of templates in decoding methods.

---

**Algorithm 1** Rotation template embedding method

---

1: Select a direction other than the direction the watermark is inserted (direction 1).
2: Rotate the template by the difference between the selected direction and direction 1. For example, if direction 9 shown in Fig. 2 (a) is selected, then direction 1 and direction 9 are 90° apart, so the template is rotated 90°.
3: Insert the rotated template in the selected direction.

---

## C. Detection Method

Figure 5 (b) shows the watermark detection process. The curvelet transformation is applied to the test image. Then the watermark pattern is generated and correlated with the curvelet image. When the correlation exceeds some pre-defined threshold value, it is determined that the watermark is detected. The correlation is expressed as

$$Correlation = \frac{\boldsymbol{C'} \cdot \boldsymbol{W}}{L} = \frac{1}{L} \sum_{m=1}^{i} \sum_{n=1}^{j} C(m,n) W(m,n), \quad (5)$$

where the notation is the same as the embedding process, and $L$ is the image size ($i \times j$). Since curvelet coefficients are robust to signal processing attacks, the watermark can be detected after such attacks as noise addition and compression.

However, it is difficult to detect the watermark after geometric distortion, because the curvelet coefficients are significantly damaged. For this case, the problem can be solved by an extraction method based on the absolute value of the curvelet coefficients, which are robust to geometric attack. The most common geometric transformations, scaling and rotation, translate and rotate the embedded watermark, respectively, as shown in Fig. 6. When the image is scaled small, high frequencies are removed, and the watermark spans scale 3 and 4, as shown in Fig. 6 (b). If the image is rotated, the frequencies rotate together, so the watermark spans direction 1 and 2, as shown in Fig. 6 (c).

If the image (and hence the watermark) has undergone a scaling attack, the effects are is similar to translating an undistorted watermark in the $\bar{F}$ domain, as shown in Figs. 6 (d) and (e). Since the watermark is inserted in the $\bar{C}$ domain, and $\bar{F}$ is the DFT of $\bar{C}$, DFT translation invariance can be exploited. Thus, even if the coefficients are translated in the $\bar{F}$ domain, coefficient magnitudes in the $\bar{C}$ domain are invariant. Therefore, if the absolute value of the curvelet is applied to Eq. 5, the watermark can be detected even after scaling attack. Since the image signal and the watermark signal are complex in the $\bar{C}$ domain, the embedded absolute value of watermark $W_{abs}$ is

$$W_{abs} = |C + W| - |C|, \qquad (6)$$

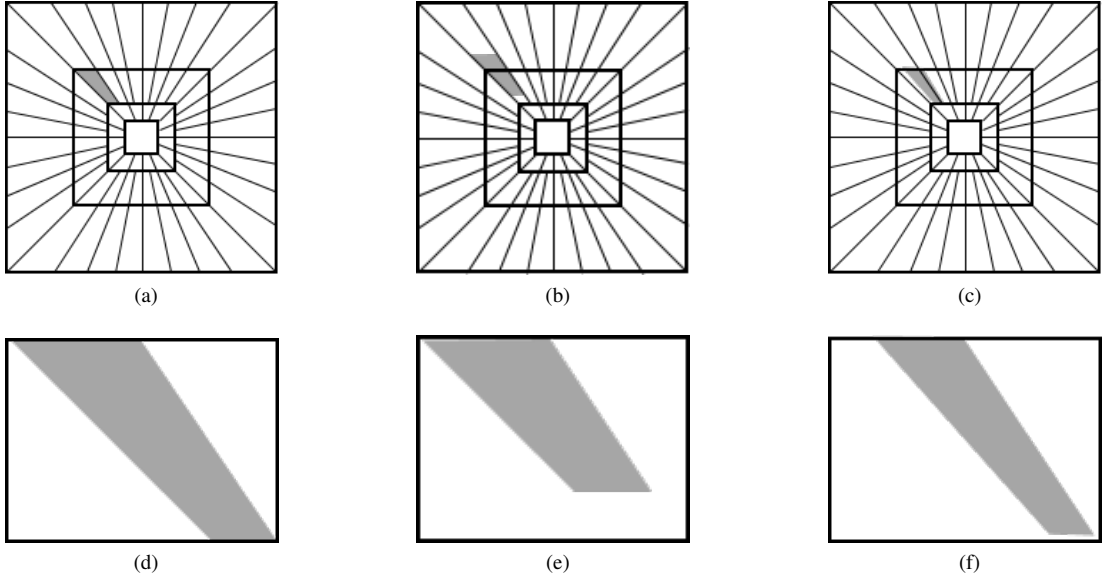

Fig. 6: Different embedded watermark positions by image scaling and rotation; T domain: (a) no attack, (b) scaling, (c) rotation; F domain: (d) no attack, (e) scaling, (f) rotation.

However, for blind detection, the original $C$ is not available, and hence $C$ and $|C|$ are not known in Eq. 6. Therefore, $W_{abs}$ can be estimated as

$$W_{abs} \simeq \tilde{W}_{abs} = |\overrightarrow{C''}| - |\overrightarrow{C'}| = |\overrightarrow{C'} + \overrightarrow{W}| - |\overrightarrow{C} + \overrightarrow{W}| = |\overrightarrow{C} + 2\overrightarrow{W}| - |\overrightarrow{C} + \overrightarrow{W}| \qquad (7)$$

where $\overrightarrow{C'} = \overrightarrow{C} + \overrightarrow{W}$ and $\overrightarrow{C''} = \overrightarrow{C'} + \overrightarrow{W}$. Figure 7 shows the vectors and absolute values. Since $C'$ and $C''$ can be obtained in the detection step, $W_{abs}$ can be estimated. The estimated absolute value of the watermark is $0 \leq \tilde{W}_{abs} \leq W_{abs}$ because the direction of $\overrightarrow{C}$ is distorted by the geometric attack. However, the error due to estimation is within the allowable range, and the watermark can be detected robustly against scaling attack.

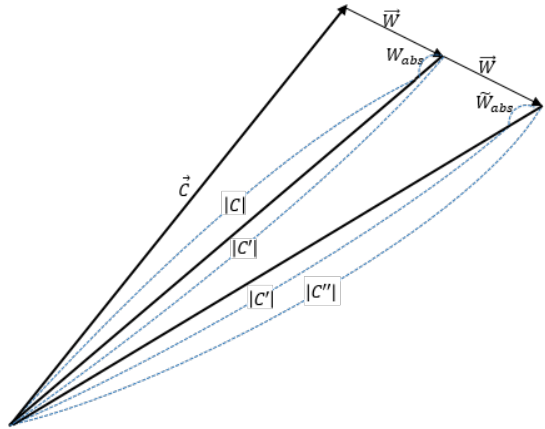

Fig. 7: Estimating the absolute value of the embedded watermark.

**Algorithm 2** Rotation template decoding method

1: Pairing directions. If the template is inserted with difference of 8, the paired directions are (1, 9), (2, 10), (3, 11), ....
2: Inversely rotate the second direction of the pair by the difference between the first and second direction of pair. This is the inverse step of Step 2 in Algorithm 1.
3: Obtain correlations for all pairs and find the pair with highest correlation.
4: Rotate the image using information from that pair. For example, if the pair found in step 3 is (3, 11), inverse rotate the image by 360°/$n_s$×2.
5: Extract the watermark using $\tilde{W}_{abs}$ from the inverse rotated image.

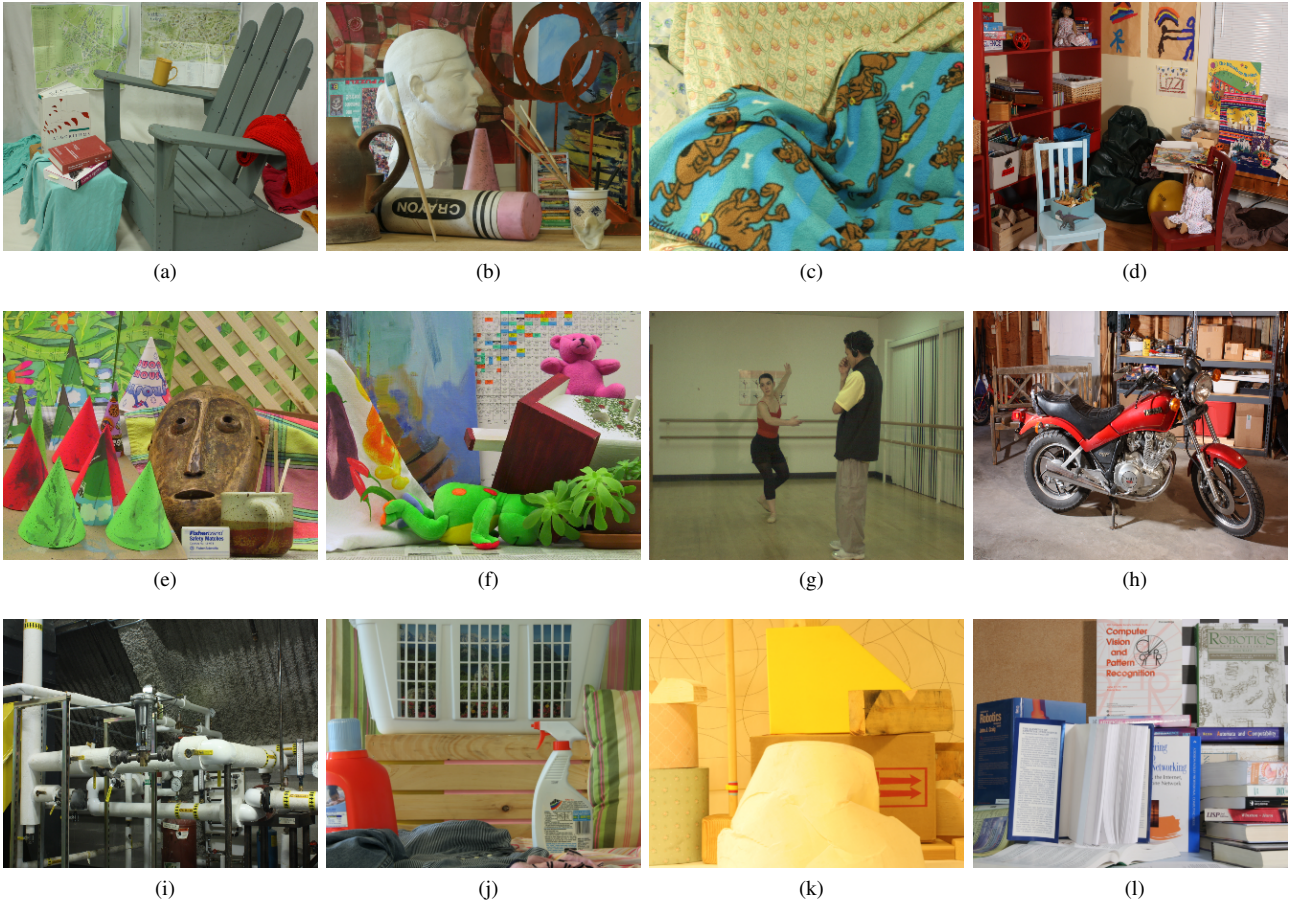

Fig. 8: Example test sets: (a) Adirondack, (b) Art, (c) Cloth, (d) Playroom, (e) Cones, (f) Teddy, (g) Ballet, (h) Motorcycle, (i) Pipes, (j) Laundry, (k) Lampshade, (l) Books.

## IV. EXPERIMENTAL RESULTS

This section shows the proposed method's invisibility and robustness to various attacks. Test image sets were obtained from Heinrich Hertz Institute [14], Microsoft Research 3D Video Datasets [15] and Middlebury [16]–[18]. The test sets consisted of approximately 800 images with resolutions from 720×576 to 1800×1500. Figure 8 shows some typical example images. We then compared the proposed method with Tao's [4] and Zhang's [2] blind curvelet domain watermarking techniques.

Tao's method is a zero-bit watermarking method using a spread spectrum, and the watermark is inserted into only one wedge. For fair comparison, the proposed method also inserted a watermark into only in one wedge and we have labelled these results as Proposed-c. In both methods, the watermark was inserted into the first wedge among 32 wedges of scale 3, and the template for the proposed method was inserted into the 9th wedge.

Zhang's method is a multi-bit watermark using a QIM method, inserting one bit per wedge, using six wedges to insert a total of six bits. For fair comparison, the proposed method also inserted watermarks in six wedges and we have labelled these results as proposed-m. In both methods, the watermark was inserted into wedges 1, 2, 3, 6, 7, and 8 among the 32 wedges of scale 3, and the template for the proposed method was inserted into the 9th wedge. In the proposed method, a direct message coding method [19] was used to insert and detect bits using the spread spectrum method.

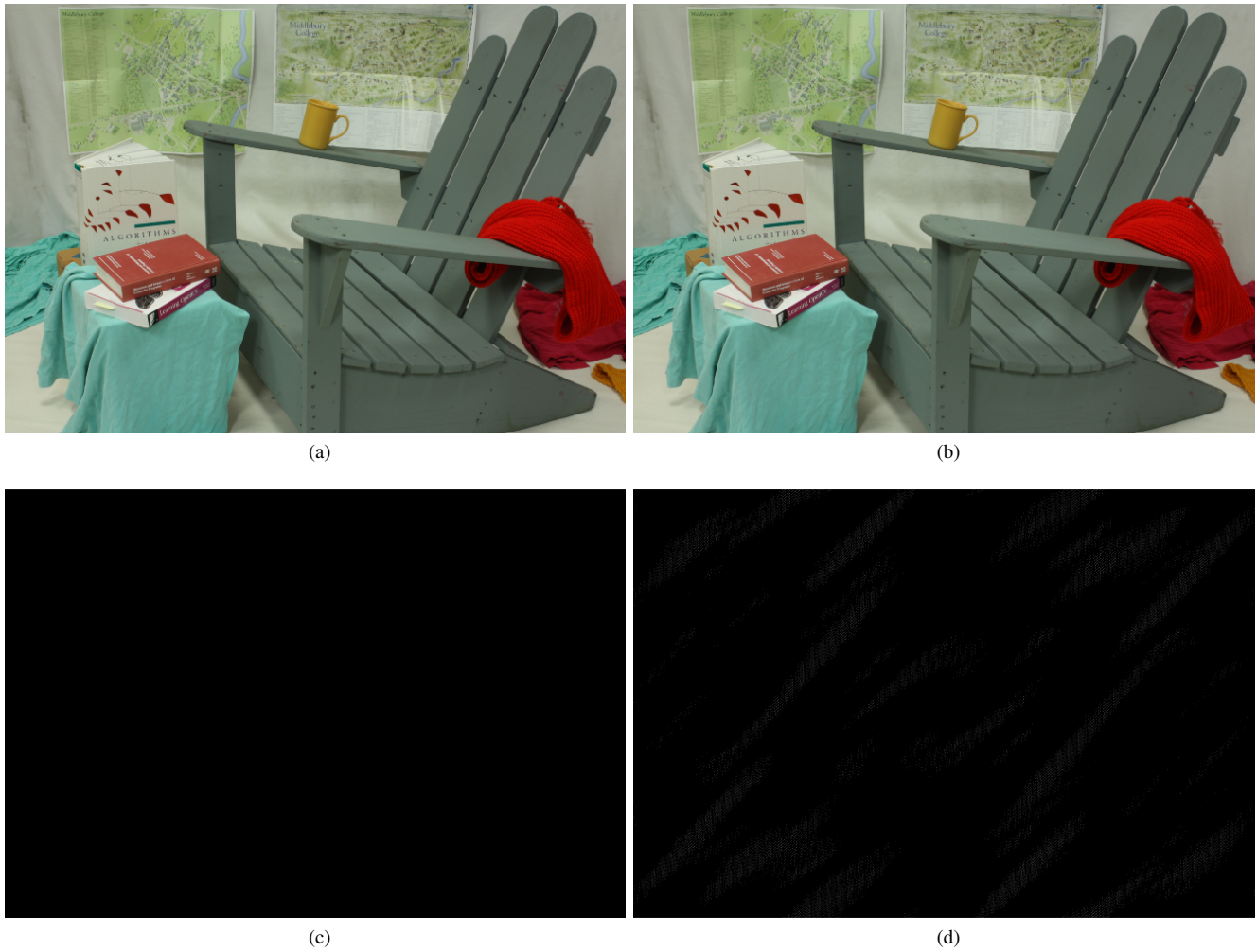

Fig. 9: Original and watermarked images: (a) original image, (b) watermarked image, (c) subtraction of original and watermarked images, (d) contrast enhanced subtraction image.

TABLE II: Average MOS

|      | Proposed-c | Tao | Proposed-m | Zhang |
|------|------------|-----|------------|-------|
| MOS  | 4.9        | 4.8 | 4.6        | 4.4   |

### A. Invisibility Test

Figures 9 (a) and (b) show typical original and watermarked images. The quality difference can hardly be distinguished by eye. Figure 9 (c) shows the difference between the watermarked and original image and Fig. 9 (d) applies 50× the contrast to Fig. 9 (c). The maximum pixel intensity difference between the watermarked and original image was only 2, which is unnoticeable without increasing the contrast. We also tested invisibility subjectively and objectively. Subjective assessments were measured by mean opinion score (MOS), and objective assessments were measured by peak signal to noise ratio (PSNR) and structure similarity (SSIM) [20]. MOS was measured by 10 image/watermark experts using the double stimulus continuous quality scale method (ITU-R [21]), with the experimental environment being a 49-inch UHD TV (model 49UF8570).

Table II shows that the MOS of the proposed method is superior to previous works. In particular, the Proposed-c method has near-perfect score (4.9), which means it was difficult to distinguish between the original and watermarked images. Table

TABLE III: Average PSNR and SSIM

|      | Proposed-c | Tao    | Proposed-m | Zhang  |
|------|------------|--------|------------|--------|
| PSNR | 57.65      | 56.47  | 51.76      | 49.18  |
| SSIM | 0.9984     | 0.9977 | 0.9946     | 0.9807 |

TABLE IV: Proposed-c and Tao's method robustness to histogram equalization

|             | Proposed-c | Tao  | Fake |
| ----------- | ---------- | ---- | ---- |
| Correlation | 2.95       | 1.40 | 0.09 |

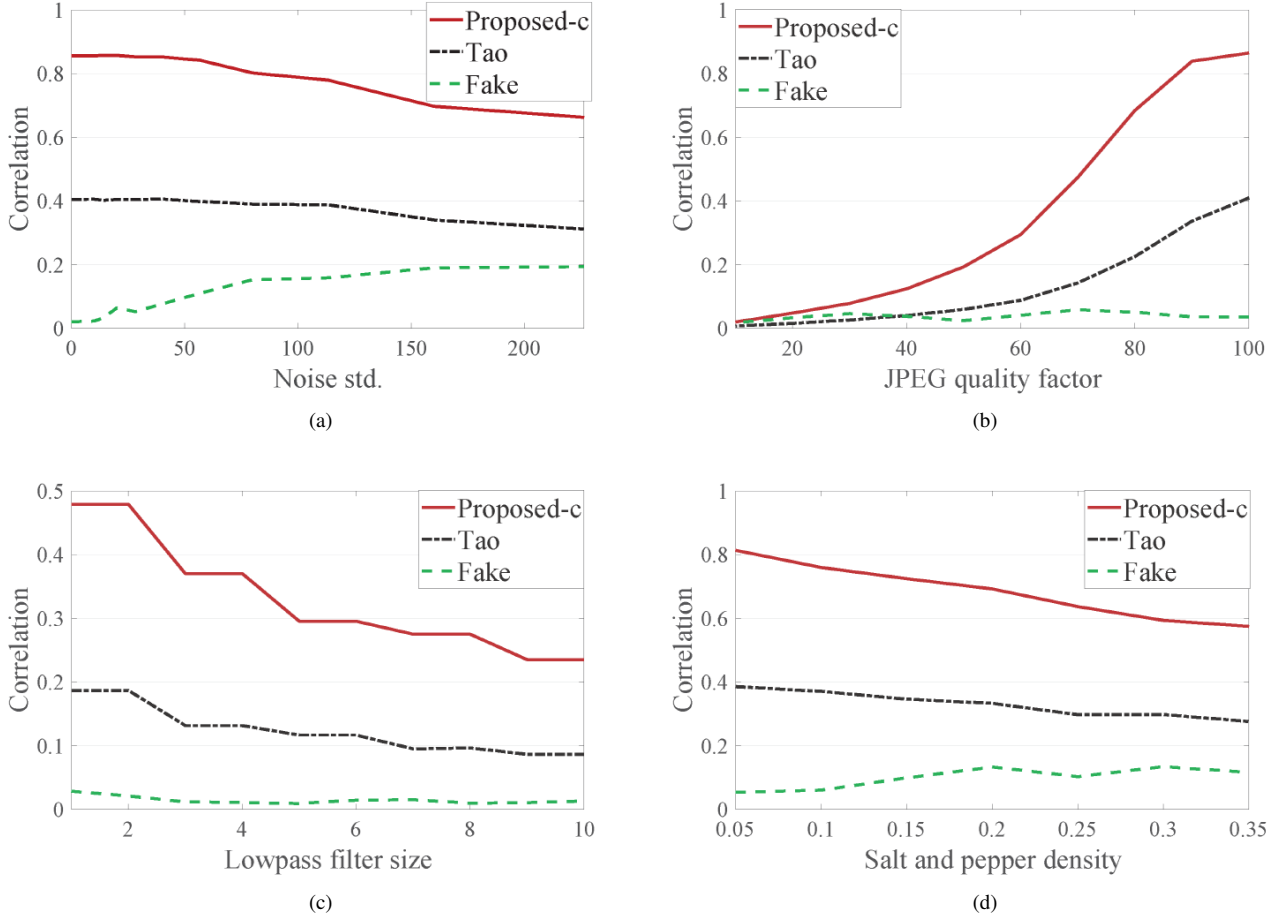

Fig. 10: Proposed-c and Tao's method Robustness to signal distortion: (a) Gaussian noise addition, (b) JPEG compression, (c) lowpass filtering, (d) salt and pepper noise addition.

III shows that for the objective assessments, PSNR and SSIM, the proposed method was more invisible than previous methods. In particular, the Proposed-c exhibited very high invisibility > 57 dB PSNR. The SSIM of Proposed-c is also the highest, so the structure of the image is best preserved. The multi-bit watermarking method proposed-m also shows better results than the Zhang's method which is the same multi-bit watermarking method in subjective and objective invisibility evaluation.

## B. Robustness to Signal Distortion

Figure 10 shows the Proposed-c and Tao's methods' robustness, which are zero-bit watermarking methods, for signal distortion. The results of Proposed-c and Tao's method showed the average correlation value between watermarked images and 'True' watermark. The 'Fake' showed the highest correlation value among the correlation between 1000 fake watermarks and watermarked images. As the results show, Proposed-c is robust to compression, filtering, and several noise addition. Correlation of Proposed-c was almost twice that of Tao's method. And Proposed-c also shows high robustness against histogram equalization, as shown in Table IV. The watermark was inserted by the same spread spectrum method, but the proposed method was more robust against signal distortion because it was not damaged by the curvelet filter.

TABLE V: Proposed-m and Zhang's method robustness to histogram equalization.

|     | Proposed-m | Zhang |
| --- | ---------- | ----- |
| BER | 0.02       | 0.38  |

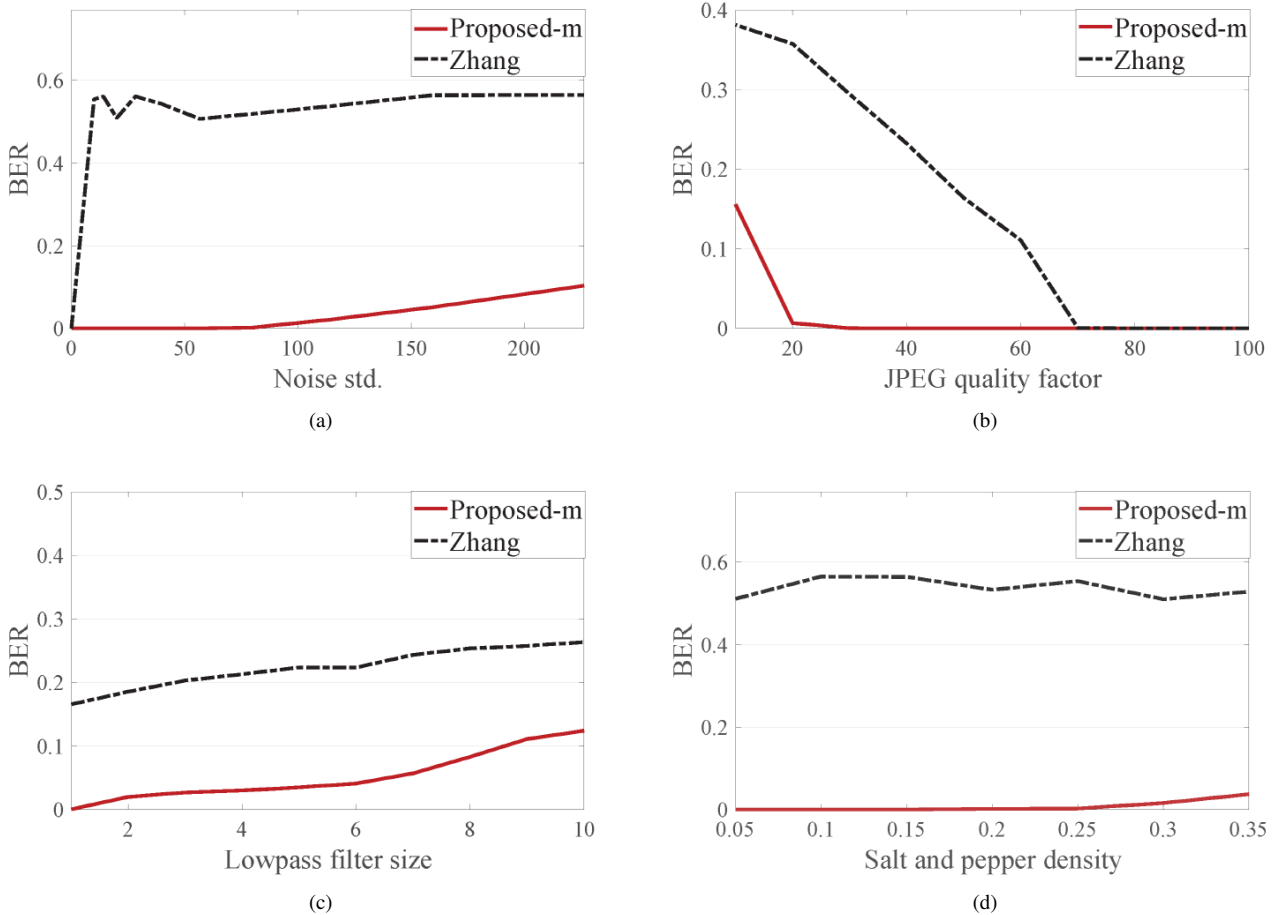

Fig. 11: Proposed-m and Zhang's method robustness to signal distortion: (a) Gaussian noise addition, (b) JPEG compression, (c) lowpass filtering, (d) salt and pepper noise addition.

Fig. 11 shows robustness of Proposed-m and Zhang's methods, which are multi-bit watermarking methods, for signal distortion. Robustness was measured using the bit error rate (BER) which is defined as,

$$BER = frac{b_e}{b_c + b_e} = frac{b_e}{b_t}, \quad (8)$$

where $b_e$ is number of error bits, $b_c$ is number of correctly decoded bits and $b_t$ is total number of decoded bits. Zhang's method exhibited significantly higher BER than the Proposed-m method. In particular, Zhang's method shows vulnerability to Gaussian and salt and pepper noise attack. This is because coefficient impairments by curvelet filtering and quantization step is relatively low compared to the noise size. As shown in Table V, Zhang's method is also vulnerable to histogram adjustments. This is because the step size of the quantized coefficients has been modified during histogram equalization. However, since there is no information on the modified step size in the decoding step, the bits can not be decoded correctly. On the other hand, the proposed method is able to detect the bits reliably even after the histogram equalization attack. This is because the correlation method is robust to histogram equalization.

### C. Robustness to Geometric Distortion

Figures 12 and 13 show Proposed-m, Tao's, and Zhang's methods' robustness to scaling and rotation. Tao's method uses a complex number of curvelet coefficients that are vulnerable to geometric attacks, and therefore it is not robust against geometric attacks. In contrast, Zhang's method exhibited high robustness to geometric attacks, since the watermark was inserted into the absolute value of the curvelet coefficients, which are less deformed in geometric attacks. Proposed-m also exhibited high robustness against geometric attacks, and would be sufficient for practical use.

Larger rotations can be addressed using the proposed template method. The template was inserted in scale 3, which is composed of 32 directions. Hence, image rotation can be detected at resolution 360°/32 = 11.25°. Figure 14 (a) shows that template accuracy is low where the template spanned two directions (e.g. 5.625°, 16.875°, 28.125°, ...). If the range of 'True' is expanded to the spanned direction, it shows high accuracy in all sections, as shown in Fig. 14 (b). After restoring the image with resolution 11.25° usisng the template, the watermark can be found through an heuristic search, which requires an acceptable amount of computation to detect the watermark.

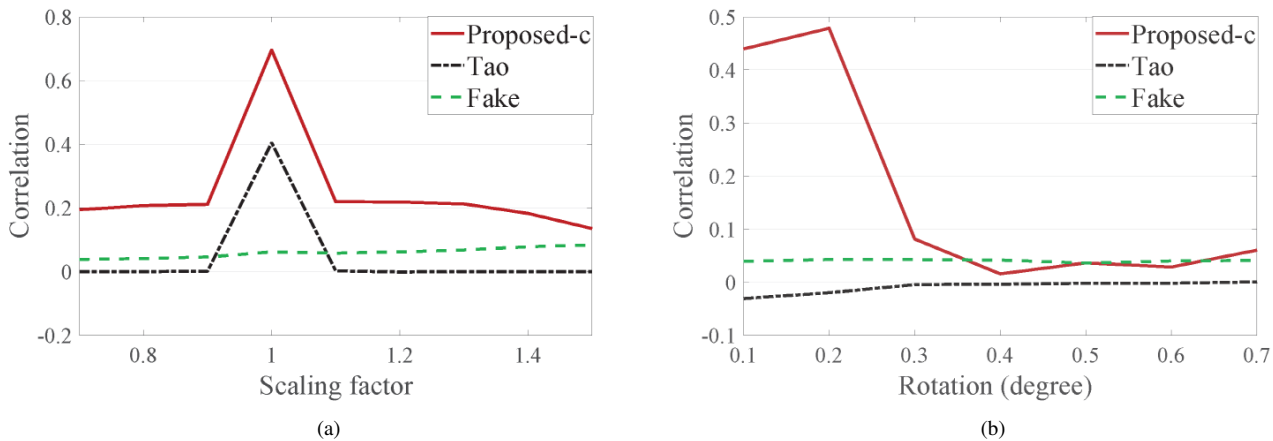

Fig. 12: Proposed-c and Tao's method robustness to geometric distortion: (a) scaling and (b) rotation.

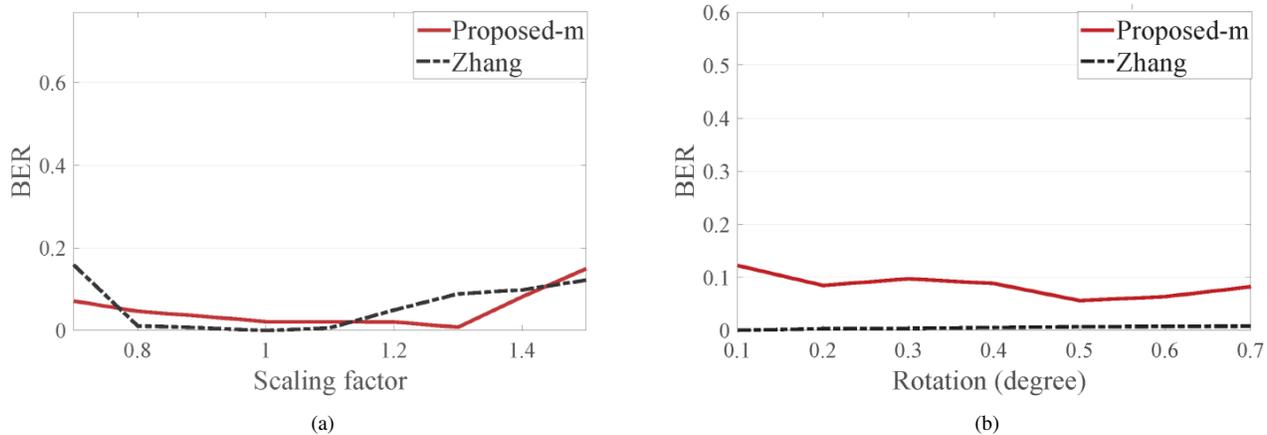

Fig. 13: Proposed-m and Zhang's method robustness to geometric distortion: (a) scaling and (b) rotation.

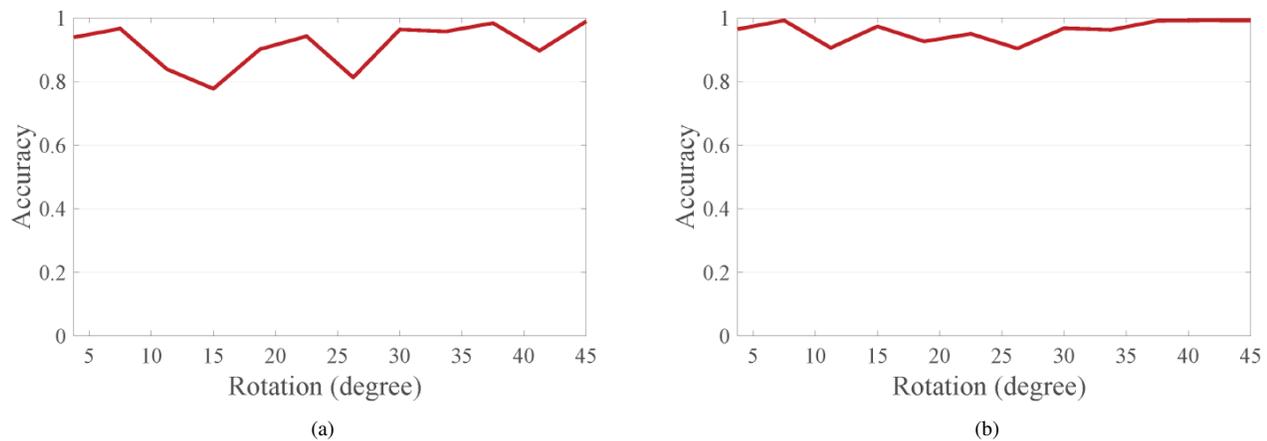

Fig. 14: Template accuracy against rotation attack: (a) True only if the template embedded direction is exactly found (b) The range of 'True' is expanded to the spanned direction.

## V. CONCLUSION

This paper proposed a blind watermarking technique based on curvelet transformation. Watermarking techniques have been widely applied to protect copyright, but quality degradation is inevitable, and many people are reluctant to embed watermarks. To overcome these shortcomings of watermarking, the proposed watermarking method minimizes quality degradation, and maximizes invisibility by using the curvelet domain while maintaining robustness against various attacks. With watermark generation technique suitable for curvelet, the proposed maximizes robustness against signal processing attack with small watermarking energy, and robustness against scaling and rotation was obtained by a template and watermark detection method using the absolute value of curvelet coefficients. Experimental results showed that the proposed method's invisibility was superior to previous methods, and robustness against signal and geometric attacks was reliable, and suitable for real-world application. Future study will expand this research into video content, minimizing video quality degradation due embedding watermarks while maintain robustness to video compression and various other attacks that occur in the video environment.